\documentclass[aps,prb,twocolumn,showpacs]{revtex4}
\usepackage{tabularx,graphicx}
\usepackage{epsfig}
\newcommand{\nin}{\noindent}
\newcommand{\vscro}{v_{\rm scr} ( 0 , \bqpar  , \bqper , \omega ) }

\newcommand{\bq}{{\bf q}}
\newcommand{\bqpar}{{\bf q}_{\parallel}}
\newcommand{\bqper}{{\bf q}_{\perp}}

\newcommand{\chiw}{\chi_{\rm w}}
\newcommand{\chiG}{\chi_{\rm G}}
\newcommand{\dw}{\nu_{\rm w}}
\newcommand{\dG}{\nu_{\rm G}}
\newcommand{\Dw}{{\cal D}_{\rm w}}
\newcommand{\DG}{{\cal D}_{\rm G}}
\newcommand{\kFw}{k_{\rm F}^{\rm w}}
\newcommand{\kFG}{k_{\rm F}^{\rm G}}
\newcommand{\lw}{l_{\rm w}}
\newcommand{\lG}{l_{\rm G}}
\newcommand{\vscrz}{v_{\rm scr} ( z , \bqpar , \bqper , \omega ) }
\newcommand{\vscrzz}{v_{\rm scr} ( \bar{z} , \bqpar , \bqper , \omega ) }
\newcommand{\eqq}{e^{- | \bq | z}}
\newcommand{\qc}{q_{\rm c}}

\begin{document}

\author{ Francisco Guinea}
\title{Electronic dephasing in wires due to metallic gates.}
\affiliation{ Instituto de Ciencia de Materiales de Madrid, CSIC,
Cantoblanco, E-28043 Madrid, Spain.}

\begin{abstract}
The dephasing effect of metallic gates on electrons moving in one
quasi--one--dimensional diffusive wires is analyzed. The
incomplete screening in this geometry implies that the effect of
the gate can be described, at high energies or temperatures, as an
electric field fluctuating in time. The resulting system can be
considered a realization of the Caldeira-Leggett model of an
environment coupled to many particles. Within the range of
temperatures where this approximation is valid, a simple
estimation of the inverse dephasing time gives $\tau_{\rm G}^{-1}
\propto T^{1/2}$.

\end{abstract}
  \maketitle
\section{Introduction.}
The low temperature dephasing time of electrons in diffusive
metals has attracted a great deal of
attention\cite{MJW97,PB02,MW03,Petal03}. Different mechanisms has
been proposed to explain the anomalous dephasing properties. Some
of them are extrinsic, like dynamical defects\cite{IFS99}, two
level systems\cite{ZDR99}, or magnetic impurities\cite{KG01}.
Alternatively, intrinsic effects have also been
proposed\cite{GZ98}. Screening effects in a
quasi--one--dimensional geometry are significantly reduced,
leading to a breakdown of Fermi liquid theory at low
temperatures\cite{AAK82,SAI90} (see also\cite{MA04}).  In the
following we study the dephasing induced by metallic gates on
quasi--one--dimensional diffusive wires.The study follows the
analysis in\cite{GJS04}, where dephasing effects in ballistic
quantum dots was considered.

As discussed in more detail below, the presence of the gate
implies the existence of two regimes: i) For distances $L \ll z$,
or time scales larger than $ \Dw^{-1} z^2$, where $\Dw$ is the
diffusion coefficient of the wire, and $z$ is the distance to the
gate, the dephasing time is the sum of a contribution from charge
fluctuations within the wire and another due to the fluctuations
at the gate. The fluctuating potential induced by the gate varies
at scales comparable or larger than $z$. Hence, the gate potential
at these scales can be calculated within the dipolar
approximation. Because of the one dimensional geometry of the
wire, this potential is not screened by the charge fluctuations of
the wire. As discussed in\cite{GJS04}, this coupling can be
considered a generalization to a many particle system of the
Caldeira-Leggett model of ohmic dissipation\cite{CL81}. This model
shows anomalous dephasing in many
situations\cite{GZ98b,IJR01,G02,G03}. ii) At distances $L \gg z$
or time scales lower than $ \Dw^{-1} z^2$ the distance between the
wire and the gate can be neglected. The screening by the gate
leads to an effective short range potential along the wire. This
effect, however, only includes logarithmic corrections in the
standard expression for the inverse dephasing time in
wires\cite{AAK82,SAI90,MA04} (see below). These different regimes
of the model are discussed in section III. The next section
generalizes the results to gates with different geometries.
Finally, section V contains a discussion of the most relevant
results. The units are such that $\hbar = 1$.
\section{The model.}
\begin{figure}[!]
  \begin{center}
       \epsfig{file=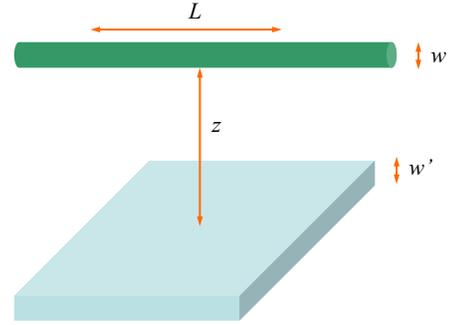,height=6cm}
    \caption{Sketch of the system considered in the text.}
    \label{wires}
\end{center}
\end{figure}

We study the setup sketched in Fig.[\ref{wires}]. A wire of width
$w$ is located at height $z$ over a two dimensional metallic gate
of width $w'$. The effects of the finite width of both systems is
included through the densities of states, $\dw$ and $\dG$, defined
as number of states per unit length and per unit area
respectively. We study the contribution to dephasing from the gate
using the scheme proposed in\cite{MA04}. The probability of
transition of a particle at the Fermi level to other states, after
time $t$, using second order perturbation theory, at temperature
$T = \beta^{-1}$, is:
\begin{widetext}
\begin{equation}
{\cal P}^{(2)} ( t ) \simeq \int_0^t d \tau \int_0^t d \tau' \int
d \bqpar \int_{| \omega | > 1/t} d \omega e^{i {\bqpar } [ {\bf r}
( \tau ) - {\bf r} ( \tau' )] + i \omega ( \tau - \tau' )} \frac{|
\omega |}{1 - e^{- \beta \omega} } {\rm Im} \left[ v_{\rm scr} (
\bq , \omega ) \right] \label{Fermi}
\end{equation}
\end{widetext}
\nin where $\bqpar$ is the momentum in the direction parallel to
the wire, and $v_{\rm scr} ( \bqpar , \omega )$ is the screened
potential between points within the wire. In general, we can write
the screened potential as $\vscrzz$, where $\bar{z}$ is the
vertical coordinate. We assume that the wire is at $\bar{z} = z$,
and the gate is at $\bar{z}=0$. The screened potential at these
two vertical coordinates can be written as:
\begin{widetext}
\begin{eqnarray} \vscrz &= &v_0 ( z , \bqpar , \bqper , \omega ) +
\frac{2 \pi e^2}{| \bq |} \chiw ( \bqpar , \omega ) \int d \bqper
\vscrz + \nonumber \\ &+ &\frac{2 \pi e^2 \eqq}{| \bq |} \chiG (
\bqpar , \bqper , \omega ) \vscro \nonumber \\ \vscro &= &v_0 ( 0
, \bqpar , \bqper , \omega ) + \frac{2 \pi e^2 \eqq}{| \bq |}
\chiw ( \bqpar , \omega ) \int d \bqper \vscrz + \nonumber \\ &+
&\frac{2 \pi e^2}{| \bq |} \chiG ( \bqpar , \bqper , \omega )
\vscro \label{vscr}
\end{eqnarray} where $\bqpar$ and $\bqper$ are the momenta parallel
and perpendicular to he wire, and $\chiw$ and $\chiG$ are the
polarizabilities of the wire and the gate. They are given by:
\begin{eqnarray}
\chiw ( \bqpar , \omega ) &= &- \frac{\dw \Dw \bqpar^2}{i \omega +
\Dw \bqpar^2} \nonumber \\ \chiG ( \bq , \omega ) &= &- \frac{ \dG
\DG \bq^2}{i \omega + \DG \bq^2} \label{chi} \end{eqnarray} where
$\dw , \dG , \Dw$ and $\DG$ are the densities of states and
diffusion coefficients of the wire and the gate, respectively.

From the second of the equations in (\ref{vscr}), we can write:
\begin{equation} \vscro = \frac{v_0 ( 0 , \bqpar , \bqper , \omega ) +
\frac{2 \pi e^2}{| \bq |} \chiw ( \bqpar , \omega ) \int d \bqper
\vscrz}{1 - \frac{2 \pi e^2}{| \bq |} \chiG ( \bqpar , \bqper ,
\omega )} \label{aux}
\end{equation} For a point charge at a point in the wire, we have:
\begin{eqnarray}
v_0 ( z , \bqpar , \bqper , \omega ) &= &\frac{2 \pi e^2}{| \bq |}
\nonumber \\ v_0 ( 0 , \bqpar , \bqper , \omega ) &= &\frac{2 \pi
e^2 \eqq}{| \bq |} \end{eqnarray} Using this expression, and
inserting eq.(\ref{aux}) into the first equation in (\ref{vscr}),
we find:

\begin{eqnarray}
v_{\rm scr} ( \bqpar , \omega ) &= &\frac{e^2 \left[ \log \left(
\frac{q_{\rm c}}{\bqpar} \right) + {\cal F} ( \bqpar , \omega
)\right]}{1 + e^2 \chiw ( \bqpar , \omega ) \left[ \log \left(
\frac{q_{\rm c}}{\bqpar} \right) +  {\cal F} ( \bqpar , \omega )
\right]} \nonumber
\\ {\cal F} ( \bqpar , \omega ) &= &\int d \bqper \frac{4 \pi^2 e^2 e^{-
2 | \bq | z}}{| \bq |^2} \frac{\chiG ( \bq , \omega )}{1 - \frac{2
\pi e^2}{| \bq |} \chiG ( \bq , \omega )} \label{selfcon}
\end{eqnarray}
where $\qc$ is a high momentum cutoff proportional to the inverse
of the width of the wire.

For sufficiently low momenta, $| \bq | \ll e^2 \dG$, we can write:
\begin{eqnarray}
{\cal F} ( \bqpar , \omega ) &\approx &  {\cal F}_1 ( \bqpar  )
- i \omega {\cal F}_2 ( \bqpar  ) \nonumber \\
{\cal F}_1 ( \bqpar  ) &= &\int d \bqper \frac{2 \pi^2 e^{- 2 |
\bq | z}}{| \bq |} \sim \log ( \bqpar z ) \nonumber \\ {\cal F}_2
( \bqpar  ) &\approx & \frac{1}{e^2 \dG \DG} \int d \bqper
\frac{e^{- 2 | \bq | z}}{| \bq |^2} \sim \frac{1}{e^2 \dG \DG}
\frac{1}{\bqpar} \label{expansion}
\end{eqnarray}
 For small frequencies, $\omega \leq \Dw \bqpar^2$, we
find:

\begin{equation}
{\rm Im} \left[ v_{\rm scr} ( \bqpar , \omega ) \right] \approx
\frac{\omega e^4 \dw}{\Dw \bqpar^2} \frac{\left[ \log \left(
\frac{q_{\rm c}}{\bqpar} \right) + {\cal F}_1 ( \bqpar  )
\right]^2 + \Dw \bqpar^2 {\cal F}_2 ( \bqpar )}{\left\{ 1 + \dw
e^2 \left[ \log \left( \frac{q_{\rm c}}{\bqpar} \right) + {\cal
F}_1 ( \bqpar ) \right] \right\}^2}
\label{calF}\end{equation} 

\section{Results.}

\subsection{Dipolar approximation ($\tau^{-1} ( T ) \gg  \Dw  / z^2$).}
 The integral over the time difference $\tau - \tau'$ in eq.(\ref{Fermi}) is
bounded by the inverse of the temperature, $\beta$. Both $z$ and
the value of $\langle \left[  {\bf r} ( \tau ) - {\bf r} ( \tau' )
\right]^2 \rangle$ for $\tau - \tau' \simeq \beta$ act as a lower
cutoff in the integrals over $\bqpar$, so that $| \bqpar |^{-2}
\ll {\rm Min} \left\{ \langle \left[  {\bf r} ( \tau ) - {\bf r} (
\tau' ) \right]^2 \rangle , z^2 \right\}$. When $\langle \left[
{\bf r} ( \tau ) - {\bf r} ( \tau' ) \right]^2 \rangle \ll z^2$,
we can limit the integral over $\bqpar$ to $| \bqpar | \leq
z^{-1}$ and substitute:
\begin{eqnarray}
e^{i {\bqpar} [ {\bf r} ( \tau ) - {\bf r} ( \tau' )]} &\approx &-
\left\{ {\bqpar} [ {\bf r} ( \tau ) - {\bf r} ( \tau' )]
\right\}^2 \label{dipolar_3}
\end{eqnarray}
 Within these approximations, and setting $q_c^{-1} \approx w$, eq.(\ref{calF}) becomes:
\begin{eqnarray}
{\rm Im} \left[ v_{\rm scr} ( \bqpar , \omega ) \right] e^{i
\bqpar [ {\bf r} ( \tau ) - {\bf r} ( \tau ' )]} &\approx &v_{\rm
1D} ( \bqpar , \omega )  e^{i \bqpar [ {\bf r} ( \tau ) - {\bf r}
( \tau ' )]} + v_{\rm 2D} ( \bqpar , \omega ) \nonumber \\ v_{\rm
1D} ( \bqpar , \omega ) &= &\frac{\omega }{\dw \Dw  \bqpar^2 } \nonumber \\
v_{\rm 2D} ( \bqpar , \omega )&= &\frac{\omega | \bqpar | [ {\bf
r} ( \tau ) - {\bf r} ( \tau' )]^2}{{e^4 \dw^2 \dG \DG } \log^2
\left( z / w \right) }\label{dipolar_2}
\end{eqnarray}
\end{widetext}
Inserting eq.(\ref{dipolar_2}) in eq.(\ref{Fermi}) one obtains
that  ${\cal P}^{(2)} ( t )$ can be written as the sum of two
contributions, ${\cal P}^{(2)}_{\rm w} ( t )$ and ${\cal
P}^{(2)}_{\rm G} ( t )$ arising from $v_{\rm 1D}$ and $v_{\rm 2D}$
(note that screening effects from the gate are included in the the
denominator of $v_{\rm 1D}$ in eq.(\ref{dipolar}) through the term
$\log ( \bq_{\rm c} z )$ ). The first term when inserted in
eq.(\ref{Fermi}), give the contribution to the function ${\cal
P}^{(2)}_{\rm G} ( t )$ calculated in\cite{MA04}, leading to the
standard expression for the dephasing in one dimensional wires.

The contribution from $v_{\rm 2D}$ in eq.(\ref{dipolar_2})to
eq.(\ref{Fermi}) can be written as:

\begin{widetext}
\begin{equation}
{\cal P}^{(2)}_{\rm G} ( t ) \simeq \frac{T}{e^4 \dw^2 \dG \DG z^2
\log^2 ( z / w )} \int_0^t d \tau \int_0^t d \tau' \int_{1/t}^{T}
d \omega \left([ {\bf r} ( \tau ) - {\bf r} ( \tau' )] \right)^2
e^{i \omega ( \tau - \tau' )} \label{dipolar}
\end{equation}
\end{widetext}
Using $\langle [  {\bf r} ( \tau ) - {\bf r} ( \tau' ) ]^2 \rangle
= \Dw ( \tau - \tau' )$, we finally obtain:
\begin{equation}
{\cal P}^{(2)}_{\rm G} ( t )  \simeq \frac{T \Dw t^2}{e^4 \dw^2
\dG \DG \log^2 ( z / w ) z^2} \label{dipolar_2}
\end{equation}
From this equation we can define a dephasing time due to the
presence of the gate, ${\cal P}^{(2)}_{\rm G} (\tau_{\rm G})
\approx 1$, as:
\begin{equation}
\hbar \tau_{\rm G}^{-1} \simeq \sqrt{\frac{T \Dw}{e^4 \dw^2 \dG
\DG \log^2 ( z / w ) z^2}}\label{t_gate}
\end{equation} On the other hand, the dephasing time due to
intrinsic processes can be written as (see also\cite{B96}):
\begin{equation}
\hbar \tau_{\rm w}^{-1} \simeq \frac{T^{2/3}}{\Dw^{1/3} \dw^{2/3}
} \label{t_int}
\end{equation} Because of the different temperature dependence,
$\tau_{\rm G}^{-1}$ is greater than $\tau_{\rm w}^{-1}$ at
temperatures below a value $T'$ given by:
\begin{equation}
T' \simeq \frac{\Dw^5}{\DG^3 \dG^3 e^{12} \dw^2 \log^6 ( z / w )
z^6} \label{T1}\end{equation} The approximations leading to this
result are valid provided that $\tau_{\rm G}^{-1} \geq \Dw / z^2$.
This condition breaks down below a temperature $T''$ given by:
\begin{equation}
T'' \simeq \frac{\Dw}{z^2} e^4 \dw^2 \dG \DG \log^2 ( z / w
)\label{T2}\end{equation} The dephasing due to the gate dominates
if a temperature range $T'' \leq T \leq T'$. The inequality $T''
\leq T'$ implies:
\begin{equation}
1 \leq f = \frac{\Dw}{e^4 \dw \dG \DG z \log^2 ( z / w
)}\label{constraint}
\end{equation}
We also have:
\begin{eqnarray}
e^2 \dw &\approx & r_{\rm s w} ( \kFw w_{\rm w} )^2 \nonumber \\
\frac{\Dw }{e^2 \dG \DG z} &\approx & \frac{1}{r_{\rm s w} ( \kFG
\lG )} \left( \frac{\lw}{z} \right)
\end{eqnarray}
where $\kFw$ and $\kFG$ are the Fermi wavevectors at the wire and
gate, $\lw$ and $\lG$ are the elastic mean free paths in the wire
and gate, $w_{\rm w}$ is the width of the wire, and $r_{\rm s w}
\sim ( e^2 \kFw ) / [ ( \hbar^2 {\kFw}^2 ) / m ]$ is inversely
proportional to the electronic density in the wire. In the
presence of a dielectric between the wire and the gate with
dielectric constant $\epsilon_0$, one has to replace the electric
charge $e^2$ by $e^2 / \epsilon_0$ in all expressions. Hence,
eq.(\ref{constraint}) can only be satisfied for very clean wires,
such that $z \ll \lw$, or in the presence of a large dielectric
constant, $\epsilon_0$.

\subsection{Low temperature regime ($\tau^{-1} ( T ) \ll \Dw  / z^2$).}.
At low temperatures the electrons diffuse coherently over
distances much larger than $z$. The dipolar approximation,
eq.(\ref{dipolar_3}) cannot be made, and the cutoff in the
integrals over $\bqpar$ is $[ \Dw ( \tau - \tau' )]^{-1}$. Then,
using eq.(\ref{calF}) we obtain,
\begin{equation}
{\cal P}^{(2)}_{\rm G} ( t ) \simeq \frac{T t}{e^4 \dw^2 \dG \DG
\log^2 ( \qc z )} \log \left( \frac{\Dw}{z^2 t} \right)
\end{equation}
Neglecting logarithmic corrections, this result leads to:
\begin{equation}
\hbar \tau_{\rm G}^{-1} \simeq \frac{T}{e^4 \dw^2 \dG \DG \log^2 (
\qc z )} \label{t_tot}
\end{equation}
In this regime there is also a contribution from the intrinsic
processes, given by eq.(\ref{t_int}). These processes will
dominate at sufficiently low temperatures, $T \leq T'''$ where:
\begin{equation}
T''' = \frac{e^{12} \dw^4 \dG^3 \DG^3 \log^6 ( \qc z )}{\Dw}
\label{T3}\end{equation} Note that when $T'' \leq T'''$, where
$T''$ is given in eq.(\ref{T2}) the contributions from the gate
are always smaller than those coming from fluctuations in the
wire. The condition $T''' \leq T''$ reduces to
eq.(\ref{constraint}).

Combining the results in this subsection and in the preceding one,
we can write:
\begin{eqnarray}
{\cal P}^{(2)}_{\rm G} ( t ) &\approx &\left\{ \begin{array}{lr}
\frac{T}{T''} \left( \frac{\Dw t}{z^2} \right)^2 &t^{-1} \ll
\frac{\Dw}{z^2} \\ \frac{T}{T''} \frac{\Dw t}{z^2}  &t^{-1} \gg
\frac{\Dw}{z^2} \end{array} \right. \nonumber \\ {\cal
P}^{(2)}_{\rm int} ( t ) &\approx &f^{-1} \frac{T}{T''} \left(
\frac{\Dw t}{ z^2} \right)^{3/2}
\end{eqnarray}
where $f$ is defined in eq.(\ref{constraint}). These expressions
lead to:
\begin{eqnarray}
\tau_{\rm G}^{-1} &\approx &\left\{ \begin{array}{lr}
\frac{\Dw}{z^2} \left( \frac{T}{T''} \right)^{1/2} &T \geq T'' \\
\frac{\Dw}{z^2} \frac{T}{T''} &T \leq T'' \end{array} \right.
\nonumber \\ \tau_{\rm int}^{-1} &\approx &\frac{\Dw}{z^2} \left(
\frac{T}{f T''} \right)^{2/3} \label{t_deph}
\end{eqnarray}
\section{Extensions to other geometries.}
\subsection{One dimensional gate.}
The analysis in the two previous sections can be extended, in a
straightforward way, to the case where the gate is another
quasi--one--dimensional wire. The gate polarizability, defined in
eq.(\ref{chi}) depends only on the momentum parallel to the wire.
The function ${\cal F}$ in eq.(\ref{selfcon}) becomes:
\begin{equation}
{\cal F} ( \bqpar , \omega ) \approx e^4 K_0^2 ( \bqpar z )
\frac{\chiG ( \bqpar , \omega )}{1 - e^2 K_0 ( \bqpar z ) \chiG (
\bqpar , \omega )}
\end{equation}
where $K_0 ( \bqpar z )$ is a modified Bessel function:
\begin{equation}
K_0 ( \bqpar z ) = \int_0^{\infty} d \bqper  \frac{\cos ( \bqper z
)}{\sqrt{\bqpar^2 + \bqper^2}}
\end{equation}

We can, as in the preceding section, study separately the dipolar
regime, $\tau^{-1} \ll \Dw / z^2$, and the long time regime,
$\tau^{-1} \gg \Dw / z^2$.  In the dipolar regime, using
eq.(\ref{dipolar_3}), we obtain:
\begin{equation}
\tau_{\rm G}^{-1} \approx \sqrt{\frac{T \dw \Dw}{e^2 \dG \DG z (
\dw + \dG )^2}}
\end{equation}
(note that now $\dw$ is a quasi--one--dimensional density of
states). The restriction $\tau^{-1} \leq \Dw / z^2$ implies that
this result is only valid for temperatures $T \geq T''$ where:
\begin{equation}
T'' = \frac{\Dw e^2 \dG \DG ( \dw + \dG )^2}{z^3 \dw}
\end{equation}
 At high temperatures, $T \geq T'$, we find $\tau_{\rm
G}^{-1} \leq \tau_{\rm int}^{-1}$, where:
\begin{equation}
T' = \frac{\Dw^5 \dw^7}{e^6 ( \dG \DG )^3 ( \dw + \dG )^6 z^3}
\end{equation}
The condition required for the relevance of the dephasing due to
the gate, $T'' \leq T'$ implies:
\begin{equation}
1 \leq \frac{\Dw \dw^2}{e^2 \dG \DG ( \dw + \dG )^2}
\label{constraint_2}
\end{equation}
It is interesting to note that this condition does not depend on
the distance between the wire and the gate.

At  temperatures below $T''$, such that $\tau^{-1} \leq \Dw /
z^2$, the combined system acts like an effective one dimensional
wire, leading to a $\tau^{-1} \propto T^{2/3}$ dependence.
\subsection{Three dimensional gate.}
We assume that quasiparticles at the gate are specularly reflected
at the boundary. Then, the screening properties of the system can
be calculated from the fluctuations of surface charges at the top
of the gate\cite{RM66}. The function ${\cal F}$ in
eq.(\ref{selfcon}) can be written as:
\begin{eqnarray}
{\cal F} ( \bqpar , \omega ) &= &\int d \bqper \frac{2 \pi e^2
e^{-2 \bq z}}{| \bq |} \frac{{\cal B} ( \bq , \omega ) - 1}{{\cal
B} ( \bq , \omega ) + 1} \nonumber \\
{\cal B} ( \bq , \omega ) &= &\frac{| \bq |}{\pi} \int d {\bq}_z
\frac{1}{( | \bq |^2 + \bq_z^2 ) \epsilon_{\rm G} ( \bq , \bq_z ,
\omega )} \nonumber \\ \epsilon_{\rm G} ( \bq , \bq_z , \omega )
&= &1 + \frac{4 \pi e^2}{| \bq |^2 + \bq_z^2} \chiG ( \bq , \bq_z
, \omega ) \label{F_3D}
\end{eqnarray}
Using this expression, the parameters $\tau_{\rm G} , T' , T''$
and $T'''$ which characterize the dephasing induced by the gate
can be calculated. One finds that they show the same dependence as
for a quasi--two--dimensional gate, with the only replacement
$\dG^{\rm 2D} \rightarrow \dG^{\rm 3D} \times z$, similarly to the
results discussed in\cite{GJS04}. Qualitatively, a three
dimensional gate behaves as a two dimensional gate of width $z$.
The constraint which needs to be satisfied for the gate induced
dephasing to be dominant is:
\begin{equation}
1 \leq  \frac{\Dw}{e^4 \dw \dG \DG z^2 \log^2 ( z / w
)}\label{constraint_3D}
\end{equation}
\subsection{Granular gate.}
We now analyze the dephasing induced by a gate made up of
disconnected metallic grains. We assume that each grain has volume
$V$ and it is characterized by a diffusion coefficient $\DG$ and
density of states $\dG$, leading to an intrinsic d. c.
conductivity $\sigma =  e^2  \dG \DG$.  Their response to an
applied field is characterized by their polarizability, ${\cal P}$
and absorption coefficient, $\gamma ( \omega ) \approx V \omega^2
/ \sigma$\cite{MW97}. The dielectric constant of the granular
system can be written as:
\begin{equation}
\epsilon ( \omega ) = V^{-1} \left( {\cal P} + \frac{i \omega
V}{e^2 \dG \DG} \right)
\end{equation}
We can insert this expression into eqs.(\ref{F_3D}), and carry out
the following steps in order to obtain the dephasing effects of
the gate. The inequality which needs to be satisfied for the gate
induced dephasing to prevail is:
\begin{equation}
1 \leq \frac{\Dw }{e^4 \dw \dG \DG z^2 ( 1 + V^{-1} P )^2}
\end{equation}
This expression, valid for grains larger than the mean free path,
is very similar to the corresponding one for a three dimensional
gate, eq.(\ref{constraint_3D}). The effects of a granular gate,
however, can be greatly enhanced by the surface roughness of the
grains\cite{AW93} (see also\cite{H86}), or for grains much smaller
than the mean free path.
\section{Conclusions.}
\begin{figure}[!]
  \begin{center}
       \epsfig{file=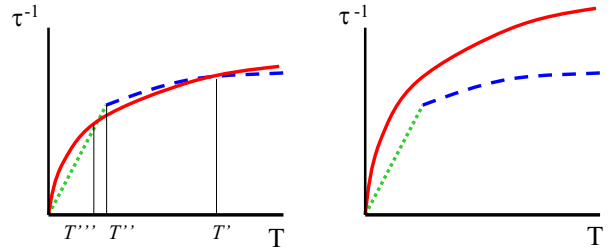,height=7cm}
    \caption{Schematic representation of the intrinsic and gate
    contributions to the inverse dephasing time,
    eq.(\protect{\ref{t_deph}}). Left: constraint
    in eq.(\protect{\ref{constraint}}) is satisfied. Full line, $\tau_{\rm
    int}$, eq.(\protect{\ref{t_int}}). Broken line, $\tau_{\rm G}$,
    eq.(\protect{\ref{t_gate}}). Dotted line, $\tau_{\rm G}$,
    eq.(\protect{\ref{t_tot}}). The values of the temperatures $T' , T''$
    and $T'''$ are given in eqs. (\protect{\ref{T1}}), (\protect{\ref{T2}})
    and (\protect{\ref{T3}}). Right: constraint
    in eq.(\protect{\ref{constraint}}) is not satisfied. The intrinsic
    contribution to the inverse dephasing time,
    eq.(\protect{\ref{t_int}}),
    dominates.}
    \label{tau_G}
\end{center}
\end{figure}
We have analyzed, within the standard approach to dephasing in
metals\cite{AAK82,SAI90,MA04} the effects of a two dimensional
diffusive gate on the coherence properties of electrons in quasi
one dimensional wires. The method used an be easily generalized to
other types of gates, like granular metals or ballistic systems.

The main results are schematically depicted in Fig.[\ref{tau_G}].
At short distances, or low temperatures, the gate induces at the
position of the wire an electric field, which fluctuates in time
but is approximately constant in space. The resulting model can be
considered an extension of the Caldeira-Leggett model\cite{CL81}
to a many particle system. Using the self consistent perturbation
theory to calculate the dephasing time, we find a $\tau_{\rm
G}^{-1} \propto T^{1/2}$ dependence. This regime requires only the
validity of the dipolar approximation, see section IIIA. Hence, it
is not restricted to the diffusive 1D wire\cite{note}. At lower
temperatures, the separation between the gate and the wire becomes
irrelevant, and the contribution from the gate changes to a
$\tau_{\rm G}^{-1} \propto T$ dependence.

The contribution to the dephasing rate due to processes intrinsic
to the wire does not change qualitatively from the standard
result\cite{AAK82,SAI90,MA04}. The screening by the gate cancels
the long range part of the electrostatic potential. The resulting
interaction, however, when treated within the RPA, leads to the
usual $\tau_{\rm int}^{-1} \propto T^{2/3}$ dependence.

The contribution from intrinsic processes dominates both at low
and high temperatures (see Fig.[\ref{tau_G}]). The existence of an
intermediate range of temperatures where the gate determines the
inverse dephasing time depends on the inequality in
eq.(\ref{constraint}). This constraint requires a mean free path
$l_{\rm w} \geq z$, which is probably not satisfied in current
experimental situations\cite{MJW97,PB02,MW03,Petal03}, where the
mean free path of the wire is comparable to its transverse
dimensions, a few nanometers. A medium between the wire and the
gate with large dielectric constant can enhance the dephasing due
to the gate. Note that a non perturbative treatment shows that the
coupling to an environment modelled by the Caldeira-Leggett model
strongly suppresses quantum coherence in a variety of
situations\cite{GZ98b,IJR01,G02,G03}.

Qualitatively similar effects can be expected for other types of
gates. If the gate is another quasi--one--dimensional wire, the
condition which determines the relative importance of the gate is
independent of the distance between the two wires,
eq.(\ref{constraint_2}).
\section{Acknowledgments.}
I am thanful to R. Jalabert and A. Zaikin for helpful comments. I
am also thankful to Ministerio de Ciencia y Tecnolog{\'\i}a
(Spain) for financial support through grant MAT2002-0495-C02-01.
\bibliography{lifetimes_wires}

\begin{thebibliography}{23}
\expandafter\ifx\csname natexlab\endcsname\relax\def\natexlab#1{#1}\fi
\expandafter\ifx\csname bibnamefont\endcsname\relax
  \def\bibnamefont#1{#1}\fi
\expandafter\ifx\csname bibfnamefont\endcsname\relax
  \def\bibfnamefont#1{#1}\fi
\expandafter\ifx\csname citenamefont\endcsname\relax
  \def\citenamefont#1{#1}\fi
\expandafter\ifx\csname url\endcsname\relax
  \def\url#1{\texttt{#1}}\fi
\expandafter\ifx\csname urlprefix\endcsname\relax\def\urlprefix{URL }\fi
\providecommand{\bibinfo}[2]{#2}
\providecommand{\eprint}[2][]{\url{#2}}

\bibitem[{\citenamefont{P.~Mohanty and Webb}(1997)}]{MJW97}
\bibinfo{author}{\bibfnamefont{E.~J.} \bibnamefont{P.~Mohanty}}
  \bibnamefont{and} \bibinfo{author}{\bibfnamefont{R.}~\bibnamefont{Webb}},
  \bibinfo{journal}{Phys. Rev. Lett.} \textbf{\bibinfo{volume}{78}},
  \bibinfo{pages}{3366} (\bibinfo{year}{1997}).

\bibitem[{\citenamefont{Pierre and Birge}(2002)}]{PB02}
\bibinfo{author}{\bibfnamefont{F.}~\bibnamefont{Pierre}} \bibnamefont{and}
  \bibinfo{author}{\bibfnamefont{N.~O.} \bibnamefont{Birge}},
  \bibinfo{journal}{Phys. Rev. Lett.} \textbf{\bibinfo{volume}{89}},
  \bibinfo{pages}{206804} (\bibinfo{year}{2002}).

\bibitem[{\citenamefont{Mohanty and Webb}(2003)}]{MW03}
\bibinfo{author}{\bibfnamefont{P.}~\bibnamefont{Mohanty}} \bibnamefont{and}
  \bibinfo{author}{\bibfnamefont{R.~A.} \bibnamefont{Webb}},
  \bibinfo{journal}{Phys. Rev. Lett.} \textbf{\bibinfo{volume}{91}},
  \bibinfo{pages}{066604} (\bibinfo{year}{2003}).

\bibitem[{\citenamefont{Pierre et~al.}(2003)\citenamefont{Pierre, Gougam,
  A.~Anthore, Esteve, and Birge}}]{Petal03}
\bibinfo{author}{\bibfnamefont{F.}~\bibnamefont{Pierre}},
  \bibinfo{author}{\bibfnamefont{A.~B.} \bibnamefont{Gougam}},
  \bibinfo{author}{\bibfnamefont{H.~P.} \bibnamefont{A.~Anthore}},
  \bibinfo{author}{\bibfnamefont{D.}~\bibnamefont{Esteve}}, \bibnamefont{and}
  \bibinfo{author}{\bibfnamefont{N.~O.} \bibnamefont{Birge}},
  \bibinfo{journal}{Phys. Rev. B} \textbf{\bibinfo{volume}{68}},
  \bibinfo{pages}{085413} (\bibinfo{year}{2003}).

\bibitem[{\citenamefont{Imry et~al.}(1999)\citenamefont{Imry, Fukuyama, and
  Schwab}}]{IFS99}
\bibinfo{author}{\bibfnamefont{Y.}~\bibnamefont{Imry}},
  \bibinfo{author}{\bibfnamefont{J.}~\bibnamefont{Fukuyama}}, \bibnamefont{and}
  \bibinfo{author}{\bibfnamefont{C.}~\bibnamefont{Schwab}},
  \bibinfo{journal}{Europhys. Lett.} \textbf{\bibinfo{volume}{47}},
  \bibinfo{pages}{608} (\bibinfo{year}{1999}).

\bibitem[{\citenamefont{Zawadovski et~al.}(1999)\citenamefont{Zawadovski, van
  Delft, and Ralph}}]{ZDR99}
\bibinfo{author}{\bibfnamefont{A.}~\bibnamefont{Zawadovski}},
  \bibinfo{author}{\bibfnamefont{J.}~\bibnamefont{van Delft}},
  \bibnamefont{and} \bibinfo{author}{\bibfnamefont{R.~C.} \bibnamefont{Ralph}},
  \bibinfo{journal}{Phys. Rev. Lett.} \textbf{\bibinfo{volume}{83}},
  \bibinfo{pages}{2632} (\bibinfo{year}{1999}).

\bibitem[{\citenamefont{Kaminski and Glazman}(2001)}]{KG01}
\bibinfo{author}{\bibfnamefont{A.}~\bibnamefont{Kaminski}} \bibnamefont{and}
  \bibinfo{author}{\bibfnamefont{L.~I.} \bibnamefont{Glazman}},
  \bibinfo{journal}{Phys. Rev. Lett.} \textbf{\bibinfo{volume}{86}},
  \bibinfo{pages}{2400} (\bibinfo{year}{2001}).

\bibitem[{\citenamefont{Golubev and Zaikin}(1998{\natexlab{a}})}]{GZ98}
\bibinfo{author}{\bibfnamefont{D.}~\bibnamefont{Golubev}} \bibnamefont{and}
  \bibinfo{author}{\bibfnamefont{A.}~\bibnamefont{Zaikin}},
  \bibinfo{journal}{Phys. Rev. Lett.} \textbf{\bibinfo{volume}{81}},
  \bibinfo{pages}{1074} (\bibinfo{year}{1998}{\natexlab{a}}).

\bibitem[{\citenamefont{Altshuler et~al.}(1982)\citenamefont{Altshuler, Aronov,
  and Khmelnitskii}}]{AAK82}
\bibinfo{author}{\bibfnamefont{B.}~\bibnamefont{Altshuler}},
  \bibinfo{author}{\bibfnamefont{A.}~\bibnamefont{Aronov}}, \bibnamefont{and}
  \bibinfo{author}{\bibfnamefont{D.}~\bibnamefont{Khmelnitskii}},
  \bibinfo{journal}{J. Phys. C} \textbf{\bibinfo{volume}{15}},
  \bibinfo{pages}{7367} (\bibinfo{year}{1982}).

\bibitem[{\citenamefont{Stern et~al.}(1990)\citenamefont{Stern, Aharonov, and
  Imry}}]{SAI90}
\bibinfo{author}{\bibfnamefont{A.}~\bibnamefont{Stern}},
  \bibinfo{author}{\bibfnamefont{Y.}~\bibnamefont{Aharonov}}, \bibnamefont{and}
  \bibinfo{author}{\bibfnamefont{Y.}~\bibnamefont{Imry}},
  \bibinfo{journal}{Phys. Rev. A} \textbf{\bibinfo{volume}{41}},
  \bibinfo{pages}{3436} (\bibinfo{year}{1990}).

\bibitem[{\citenamefont{Montambaux and Akkermans}(2004)}]{MA04}
\bibinfo{author}{\bibfnamefont{G.}~\bibnamefont{Montambaux}} \bibnamefont{and}
  \bibinfo{author}{\bibfnamefont{E.}~\bibnamefont{Akkermans}}
  (\bibinfo{year}{2004}), \eprint{cond-mat/0404361}.

\bibitem[{\citenamefont{Guinea et~al.}(2004)\citenamefont{Guinea, Jalabert, and
  Sols}}]{GJS04}
\bibinfo{author}{\bibfnamefont{F.}~\bibnamefont{Guinea}},
  \bibinfo{author}{\bibfnamefont{R.~A.} \bibnamefont{Jalabert}},
  \bibnamefont{and} \bibinfo{author}{\bibfnamefont{F.}~\bibnamefont{Sols}}
  (\bibinfo{year}{2004}), \eprint{cond-mat/0402277}.

\bibitem[{\citenamefont{Caldeira and Leggett}(1981)}]{CL81}
\bibinfo{author}{\bibfnamefont{A.~O.} \bibnamefont{Caldeira}} \bibnamefont{and}
  \bibinfo{author}{\bibfnamefont{A.}~\bibnamefont{Leggett}},
  \bibinfo{journal}{Phys. Rev. Lett.} \textbf{\bibinfo{volume}{46}},
  \bibinfo{pages}{211} (\bibinfo{year}{1981}).

\bibitem[{\citenamefont{Golubev and Zaikin}(1998{\natexlab{b}})}]{GZ98b}
\bibinfo{author}{\bibfnamefont{D.~S.} \bibnamefont{Golubev}} \bibnamefont{and}
  \bibinfo{author}{\bibfnamefont{A.}~\bibnamefont{Zaikin}},
  \bibinfo{journal}{Physica B} \textbf{\bibinfo{volume}{255}},
  \bibinfo{pages}{164} (\bibinfo{year}{1998}{\natexlab{b}}).

\bibitem[{\citenamefont{Ingold et~al.}(2001)\citenamefont{Ingold, Jalabert, and
  Richter}}]{IJR01}
\bibinfo{author}{\bibfnamefont{G.-L.} \bibnamefont{Ingold}},
  \bibinfo{author}{\bibfnamefont{R.~A.} \bibnamefont{Jalabert}},
  \bibnamefont{and} \bibinfo{author}{\bibfnamefont{K.}~\bibnamefont{Richter}},
  \bibinfo{journal}{Am. J. Phys.} \textbf{\bibinfo{volume}{69}},
  \bibinfo{pages}{201} (\bibinfo{year}{2001}).

\bibitem[{\citenamefont{Guinea}(2002)}]{G02}
\bibinfo{author}{\bibfnamefont{F.}~\bibnamefont{Guinea}},
  \bibinfo{journal}{Phys. Rev. B} \textbf{\bibinfo{volume}{65}},
  \bibinfo{pages}{205317} (\bibinfo{year}{2002}).

\bibitem[{\citenamefont{Guinea}(2003)}]{G03}
\bibinfo{author}{\bibfnamefont{F.}~\bibnamefont{Guinea}},
  \bibinfo{journal}{Phys. Rev. B} \textbf{\bibinfo{volume}{67}},
  \bibinfo{pages}{045103} (\bibinfo{year}{2003}).

\bibitem[{\citenamefont{Blanter}(1996)}]{B96}
\bibinfo{author}{\bibfnamefont{Y.~M.} \bibnamefont{Blanter}},
  \bibinfo{journal}{Phys. Rev. B} \textbf{\bibinfo{volume}{54}},
  \bibinfo{pages}{12807} (\bibinfo{year}{1996}).

\bibitem[{\citenamefont{Ritchie and Marusak}(1966)}]{RM66}
\bibinfo{author}{\bibfnamefont{R.~H.} \bibnamefont{Ritchie}} \bibnamefont{and}
  \bibinfo{author}{\bibfnamefont{A.~L.} \bibnamefont{Marusak}},
  \bibinfo{journal}{Surf. Sc.} \textbf{\bibinfo{volume}{4}},
  \bibinfo{pages}{234} (\bibinfo{year}{1966}).

\bibitem[{\citenamefont{Mehlig and Wilkinson}(1997)}]{MW97}
\bibinfo{author}{\bibfnamefont{B.}~\bibnamefont{Mehlig}} \bibnamefont{and}
  \bibinfo{author}{\bibfnamefont{M.}~\bibnamefont{Wilkinson}},
  \bibinfo{journal}{J. Phys. C: Condens. Matter} \textbf{\bibinfo{volume}{9}},
  \bibinfo{pages}{3277} (\bibinfo{year}{1997}).

\bibitem[{\citenamefont{Austin and Wilkinson}(1993)}]{AW93}
\bibinfo{author}{\bibfnamefont{E.~J.} \bibnamefont{Austin}} \bibnamefont{and}
  \bibinfo{author}{\bibfnamefont{M.}~\bibnamefont{Wilkinson}},
  \bibinfo{journal}{J. Phys. C: Condens. Matter} \textbf{\bibinfo{volume}{5}},
  \bibinfo{pages}{8461} (\bibinfo{year}{1993}).

\bibitem[{\citenamefont{Halperin}(1986)}]{H86}
\bibinfo{author}{\bibfnamefont{W.~P.} \bibnamefont{Halperin}},
  \bibinfo{journal}{Rev. Mod. Phys.} \textbf{\bibinfo{volume}{56}},
  \bibinfo{pages}{533} (\bibinfo{year}{1986}).

\bibitem[{not()}]{note}
\bibinfo{note}{It is interesting to note that, for a ballistic system, $\langle
  [ {\bf r} ( \tau ) - {\bf r} ( \tau' ) ]^2 \rangle \approx v_{\rm F}^2 ( \tau
  - \tau' )^2$. This result implies that $\tau_{\rm G}^{-1} \propto T^{1/3}$ in
  this case.}

\end{thebibliography}
\end{document}